\documentclass{vldb}

\usepackage{color}
\usepackage{epsfig}
\usepackage{multirow}
\usepackage{rotating}
\usepackage{url}
\usepackage{graphicx}
\usepackage{algorithm}
\usepackage{algpseudocode}
\usepackage{balance}
\usepackage{fixltx2e}  
\usepackage{epstopdf}
\usepackage{xcolor}
\usepackage{listings}

\newcommand{\eat}[1]{}

\newcommand{\Paragraph}[1]{\smallskip\noindent{\bf #1.}}

\newcommand{\Paragraphcolon}[1]{\smallskip\noindent{\bf #1:}}

\newcommand{\insn}[1]{{\tt\small #1}}

\begin{document}


\title{Efficient Graph Computation for Node2Vec}

\numberofauthors{1}
\author{
Dongyan Zhou\hspace{0.6in}
Songjie Niu\hspace{0.6in}
Shimin Chen\thanks{\small Corresponding author}\vspace{0.08in}\\
       \affaddr{State Key Laboratory of Computer Architecture}\\
       \affaddr{Institute of Computing Technology, Chinese Academy of Sciences}\\
       \affaddr{University of Chinese Academy of Sciences}\\
       \affaddrit{ \{zhoudongyan,niusongjie,chensm\}@ict.ac.cn}
}

\maketitle


\begin{abstract}
Node2Vec is a state-of-the-art general-purpose feature learning method
for network analysis.  However, current solutions cannot run Node2Vec
on large-scale graphs with billions of vertices and edges, which are
common in real-world applications.  The existing distributed Node2Vec
on Spark incurs significant space and time overhead.  It runs out of
memory even for mid-sized graphs with millions of vertices.  Moreover,
it considers at most 30 edges for every vertex in generating random
walks, causing poor result quality.

In this paper, we propose Fast-Node2Vec, a family of efficient
Node2Vec random walk algorithms on a Pregel-like graph computation
framework.  Fast-Node2Vec computes transition probabilities during
random walks to reduce memory space consumption and computation
overhead for large-scale graphs.  The Pregel-like scheme avoids space
and time overhead of Spark's read-only RDD structures and shuffle
operations.  Moreover, we propose a number of optimization techniques
to further reduce the computation overhead for popular vertices with
large degrees.  Empirical evaluation show that Fast-Node2Vec is
capable of computing Node2Vec on graphs with billions of vertices and
edges on a mid-sized machine cluster.  Compared to Spark-Node2Vec,
Fast-Node2Vec achieves 7.7--122x speedups.

%

\end{abstract}

\section{Introduction}
\label{sec:intro}


Graph is an important big data model, widely used to represent
real-world entities and relationships in applications ranging from the
World Wide Web~\cite{BroderKMRRSTW00}, social
networks~\cite{Liben-NowellK03}, publication
networks~\cite{WangHJTZYG10}, to protein-protein interaction networks
in bioinformatics~\cite{radivojac2013a}.  One promising approach to
network (graph)\footnote{\small In this paper, we will use
\emph{network} and \emph{graph} interchangeably.} analysis is to
construct feature vectors to represent vertices or edges in a graph
such that classical machine learning algorithms can be applied to the
resulting vector representations for network analysis tasks, such as
node classification~\cite{TsoumakasK07} and link
prediction~\cite{Liben-NowellK03}.  Node2Vec~\cite{GroverL16} is a
state-of-the-art general-purpose feature learning method for network
analysis.  It has been shown that Node2Vec achieves better accuracy
than other competitive feature learning solutions, including Spectral
Clustering~\cite{TangL11}, DeepWalk~\cite{PerozziAS14}, and
LINE~\cite{TangQWZYM15}, as well as a number of popular heuristic
solutions~\cite{GroverL16}.


Node2Vec extends the Skip-gram model~\cite{abs-1301-3781} to the
network analysis scenario.  The Skip-gram model is originally studied
in the text analysis scenario~\cite{abs-1301-3781, MikolovSCCD13}.
The goal is to automatically learn a feature vector for every word.
The first step is to sample the neighbor words for every word.  Based
on the samples, it then formulates and solves an optimization problem
using Stochastic Gradient Descent (SGD) to obtain the vector
representations of words.  In the text scenario, the neighbor words of
a word can be easily defined as the $t$ words prior to and $t$ words
following the target word in sentences, where $t$ is a parameter
(e.g., 5).  Therefore, one of the main challenges to employ this idea
to graphs is to compute the neighbors of a vertex in a graph.

The main distinctive characteristic of Node2Vec is that it proposes a
biased 2nd-order random walk model to sample neighbors of every vertex
in a graph, upon which the optimization problem similar to that in the
original text analysis scenario~\cite{MikolovSCCD13} is solved to
obtain the vertex feature vectors.  This random walk model combines
BFS and DFS in a flexible manner so that local and global structures
of the vertex neighborhood can both be captured.  In this way, it is
capable of supporting different varieties of graphs and analysis tasks
well.  The Node2Vec paper reports the run times for graphs with up to
1 million vertices and 10 million edges.   In this paper, we are
interested in efficiently supporting Node2Vec on large-scale graphs
with billions of vertices and edges, which are common in real-world
applications.


\Paragraph{Problems of Existing Node2Vec Implementations} There are
two reference implementations (i.e. Python and C++) on the Node2Vec
project page\footnote{\small\url{http://snap.stanford.edu/node2vec/}}.
Both of them run on a single machine and perform well on small graphs.
However, the data structures for large-scale graphs with billion of
vertices cannot fit into the main memory of a single machine,
demanding distributed graph computation solutions.


Spark\footnote{\small\url{https://spark.apache.org/}}~\cite{ZahariaCDDMMFSS12}
is a popular distributed computation framework for big data
processing.  Compared to the previous MapReduce
framework~\cite{DeanG04}, Spark takes advantage of in-memory
processing and caching to significantly improve computation
efficiency.  There is a Node2Vec implementation on Spark available
online\footnote{\small\url{https://github.com/aditya-grover/node2vec}}.
While it leverages Spark to compute Node2Vec on larger graphs using a
cluster of machines, there are two main problems of Spark-Node2Vec.


First, Spark-Node2Vec is not an exact Node2Vec implementation.
For every vertex, it considers at most \emph{30 edges} with the
highest weights for computing the biased random walks in order to save
memory space for storing the transition probabilities. However, a
popular vertex in a real-world graph (e.g., a social network) can have
thousands or even millions of edges.  As a result, Spark-Node2Vec
samples only a very small fraction of neighbors of a popular vertex
(e.g., 3\% of 1K neighbors, 0.003\% of 1M neighbors).  This 
significantly degrades the quality of the resulting vectors, as will be
shown in our experimental study in Section~\ref{sec:results}.
 

Second, Spark-Node2Vec runs out of memory and incurs significant
overhead for even mid-sized graphs.  Spark's core data structure is
RDD, which is a distributed data set that allows parallel computation
across multiple worker machine nodes.  To simplify data consistency
and fault tolerance, Spark's RDD is designed to be read-only.  That
is, any modification to an RDD will result in a new RDD.  However,
every random walk step needs to update the sampled walks. A typical
configuration runs 10 rounds of random walks, each with 80 steps.
Therefore, Spark-Node2Vec incurs frequent creations of new RDDs, which
consume the main memory space quickly.  Moreover, the solution employs
an RDD Join operation in selecting the transition probabilities at
every step of the random walks.  The Join operation performs data
re-distribution, a.k.a. shuffle, which spills intermediate data to
disks, incurring significant I/O overhead.


\Paragraph{Our Solution: Fast-Node2Vec on a Pregel-Like Graph
Computation Framework}  We propose a Fast-Node2Vec algorithm on a
Pregel-like graph computation framework~\cite{MalewiczABDHLC10} to
address the problems of the existing solutions.  Node2Vec consists of
a biased random walk stage and a Skip-Gram computation stage.  As the
former constitutes 98.8\% of the total execution time of
Spark-Node2Vec, we mainly focus on improving the Node2Vec random walk
stage in this paper.

First, we employ
GraphLite\footnote{\small\url{https://github.com/schencoding/GraphLite}},
an open-source C/C++ implementation of Pregel~\cite{MalewiczABDHLC10}
in order to avoid the overhead in Spark.  Pregel is a distributed
in-memory graph computation framework.  Graph computation is
implemented as programs running on individual vertices.  In contrast
to Spark, vertex states are updated in place and vertices communicate
through messages in main memory and across network.  Therefore, the
Pregel framework does not incur the RDD creation and the shuffle
overhead as in Spark. 


Second, Fast-Node2Vec computes the transition probabilities on demand
during random walks.  It does not pre-compute and store all the
transition probabilities before random walks.  As Node2Vec performs a
2nd-order random walk, the transition probability depends on a pair of
vertices in the last two steps in a random walk.  Therefore, for a
vertex $V_i$ with $d_i$ neighbors in an undirected graph, there are
$d_i^2$ transition probabilities.  The total number of transition
probabilities for all vertices is much larger than the number of
vertices and the number of edges added up together.  This is
especially the case in power-law graphs, where a small number of
popular vertices have high degrees.  We find that the total amount of
memory space required to store all transition probabilities of
large-scale graphs can be magnitudes larger than the total memory size
of the machine cluster used in our evaluation. 
Therefore, Fast-Node2Vec performs on-demand computation of transition
probabilities in order to reduce the memory space and support
large-scale graphs.

Third, we further analyze the computation overhead in Fast-Node2Vec.
We find that the communication of vertex neighbors is a significant
source of overhead, especially for popular vertices with a large
number of neighbors.  To reduce this overhead, we propose and study a
family of Fast-Node2Vec algorithms: (i) For vertices in the same graph
partition, we can directly read the neighbor information without
sending messages (FN-Local); (ii) To find the common neighbors of a
popular vertex $B$ and a low-degree vertex $S$, we may always send
$S$'s neighbors to $B$ (FN-Switch); (iii) After receiving the neighbor
information of a popular vertex $B$ from a remote graph partition, we
can cache it locally and reuse it (FN-Cache).  The above Fast-Node2Vec
algorithms are all exact implementation of Node2Vec random walks.  In
addition, we propose a variant of Fast-Node2Vec (FN-Approx) that
approximately computes next moves at popular vertices with low
overhead and bounded errors.

Finally, we perform extensive experiments to empirically evaluate all
the proposed algorithms using both real-world and synthetic data sets.
Our experimental results on a mid-sized cluster of 12 machines show
that our proposed Fast-Node2Vec solutions are capable of computing
Node2Vec for large-scale graphs with billions of vertices in a
reasonable amount of time.  Compared to Spark-Node2Vec, our proposed
Fast-Node2Vec solutions achieve 7.7--122x speedups.



\Paragraphcolon{Contributions of the Paper}  
%
\begin{list}{(\arabic{enumi})}{\setlength{\leftmargin}{5mm}\setlength{\itemindent}{0mm}\setlength{\topsep}{0.5mm}\setlength{\itemsep}{0mm}\setlength{\parsep}{0.5mm}}

    \setcounter{enumi}{0}

    \addtocounter{enumi}{1}
    \item We propose Fast-Node2Vec, a family of efficient Node2Vec
random walk algorithms on a Pregel-like graph computation framework.
Fast-Node2Vec computes transition probabilities during random walks to
reduce memory space consumption and computation overhead for
large-scale graphs.

    \addtocounter{enumi}{1}
    \item We analyze the behavior of the Node2Vec random walks and
show that popular vertices with a large number of edges incur
significant overhead in message sizes and computations.  Then we
propose and study a number of optimization techniques (including
FN-Cache and FN-Approx) to reduce the overhead.

    \addtocounter{enumi}{1}
    \item We perform extensive experiments to empirically evaluate our
proposed solutions.  In our experiments, we see that (i)
Spark-Node2Vec has both poor result quality and poor efficiency. (ii)
Fast-Node2Vec is capable of supporting graphs with billions vertices
and edges using a mid-sized machine cluster. Similar computing
resources are often available to academic researchers.  (iii) Compared
to Spark-Node2Vec, Fast-Node2Vec solutions achieve 7.7--122x speedups
speedups. (iv) As the vertex distribution of a graph becomes more and
more skewed, our optimization techniques to reduce the overhead of
popular vertices become more effective.

\end{list}

\vspace{2mm}
The rest of the paper is organized as follows.
Section~\ref{sec:graphx} briefly overviews existing Node2Vec
solutions.  Section~\ref{sec:graphlite} presents the technical details
for improving the efficiency of Node2Vec using Fast-Node2Vec on a
Pregel-Like graph computation framework. Section~\ref{sec:results}
compares Fast-Node2Vec with Spark-Node2Vec, and studies the proposed
optimizations for Fast-Node2Vec, using both real-world and synthetic
graphs in a distributed machine environment.  Finally,
Section~\ref{sec:conclusion} concludes the paper.

\section{Problems of Spark-Node2Vec}
\label{sec:graphx}


We briefly overview the Node2Vec random walk model, then discuss
Spark-Node2Vec and its problems.  

\subsection{Node2Vec Random Walk}
\label{subsec:node2vec-walk}

\begin{figure}[t]
\centering{$
  \begin{array}{c}
  \includegraphics[height=2.80cm]{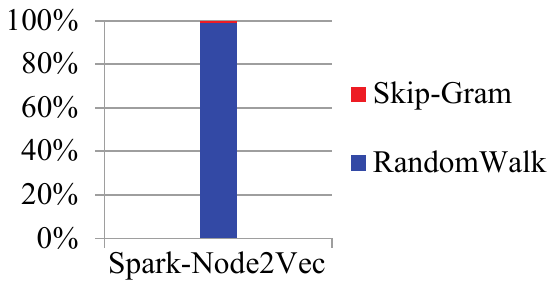}
  \end{array}
  \vspace{-0.1in}
$}
  \caption{Node2Vec runtime breakdown for the Spark implementation.}
  \label{fig:breakdown-of-runtime}
  \vspace{-0.1in}
\end{figure}

The Node2Vec~\cite{GroverL16} method performs random walks to sample
vertex neighborhood and then solves an optimization problem based on
the Skip-gram model~\cite{abs-1301-3781} to obtain vector
representations for vertices in a graph.
Figure~\ref{fig:breakdown-of-runtime} shows the runtime breakdown for
Node2Vec on the BlogCatalog graph~\cite{zafarani2009social}.  The
random walk stage takes 98.8\% of the total run time of
Spark-Node2Vec.  Moreover, the optimization problem is solved by
performing a Stochastic Gradient Descent (SGD) computation, which has
good distributed implementations.  Therefore, we focus on the
efficiency of the Node2Vec random walk stage in this paper.

\begin{figure}[h]
  \vspace{-0.05in}
\centering{$
  \begin{array}{c}
  \includegraphics[height=1in]{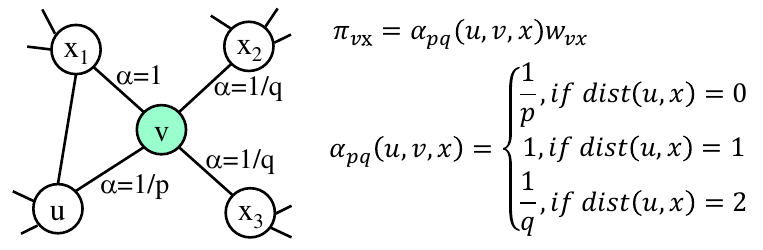}
  \end{array}
  \vspace{-0.1in}
$}
  \caption{Node2Vec performs 2nd-order biased random walks.}
  \label{fig:node2vec-random-walk}
  \vspace{-0.05in}
\end{figure}

Node2Vec simulates $r$ (e.g., $r=10$) random walks of fixed length $l$
(e.g., $l=80$) starting from every vertex.  The Node2Vec random walk
model is illustrated in Figure~\ref{fig:node2vec-random-walk}. The
unnormalized transition probability $\pi_{vw}$ from the current vertex
$v$ to its neighbor $x$ depends on the edge weight $w_{vx}$ and
$\alpha_{pq}(u,v,x)$. The latter varies based on the distance between
$x$ and the last vertex $u$ seen in the random walk.  This random walk
model flexibly combines characteristics of BFS and DFS by using the
parameters $p$ and $q$.  The lower the parameter $p$ and the higher
the parameter $q$, the more likely the walk explores the local area
close to the starting vertex $u$, mimicking a BFS-like behavior.  In
contrast, a large $p$ and a small $q$ will increase the probability
for the walk to visit vertices further away from $u$, showing a
DFS-like behavior.  In this way, the vector representations combine
both structural equivalence and local community features.  As a
result, Node2Vec can effectively support a variety of analysis
tasks~\cite{GroverL16}.

%
%

\subsection{Spark-Node2Vec}
\label{subsec:spark-node2vec}

Spark~\cite{ZahariaCDDMMFSS12} is a popular second-generation
distributed computation framework for big data processing.  It
improves upon the first-generation framework,
MapReduce~\cite{DeanG04}, by putting the results of intermediate
computation steps in memory, thereby reducing I/O overhead to access
the underlying distributed file system (e.g. HDFS) in MapReduce.  The
core data structure in Spark is RDD (Resilient Distributed Data sets).
An RDD is an immutable, distributed data set.  Operations (a.k.a.
transformations) on an RDD can be performed in parallel across all the
partitions of the RDD.  GraphX is a graph computation framework built
on top of Spark.  It represents vertices, edges, and intermediate
states as RDDs, and implements graph computations using RDD
transformations, such as joins.



Spark-Node2Vec implements random walks in two phases: 
\begin{list}{(\roman{enumi})}{\setlength{\leftmargin}{5mm}\setlength{\itemindent}{0mm}\setlength{\topsep}{0.5mm}\setlength{\itemsep}{0mm}\setlength{\parsep}{0.5mm}}

\setcounter{enumi}{0}

\addtocounter{enumi}{1}
\item \emph{Preprocessing phase}: Spark-Node2Vec pre-computes all
transition probabilities and allocates RDDs for storing the transition
probabilities and to facilitate random walks.  Every edge stores three
arrays of equal length.  One array records the neighbors of the
destination vertex.  The other two arrays are initialized using the
transition probabilities for Alias Sampling~\cite{Vose91}.  Every
vertex contains an array of (neighbor, edge weight) pairs.  To reduce
the allocated memory space and computation overhead, Spark-Node2Vec
trims the graph by removing edges.  It preserves only \emph{30 edges}
with the top most edge weights for every vertex.



\addtocounter{enumi}{1}
\item \emph{Random walk phase}: Spark-Node2Vec simulates random walks
starting from all vertices based on the pre-computed transition
probabilities.  It computes one step of all walks in every loop
iteration.  This is achieved by joining the last two walk steps, the
pre-computed transition probabilities for edges, and the recorded edge
weights for vertices.  The computed random walks are recorded in an
RDD, where the walk length grows one step per iteration for every
vertex.


\end{list}

\noindent
Spark-Node2Vec incurs both poor quality and poor efficiency:
\begin{list}{\labelitemi}{\setlength{\leftmargin}{5mm}\setlength{\itemindent}{0mm}\setlength{\topsep}{0.5mm}\setlength{\itemsep}{0mm}\setlength{\parsep}{0.5mm}}

\item \emph{Poor quality}: The trim idea over-simplifies the random
walk process.  For a vertex $v$ with $d >> 30$ outgoing edges,
Spark-Node2Vec would explore only 30 neighbors of $v$.  For $d=1000$,
this is 3\% of all neighbors.  For $d=$1 million, merely 0.003\% of
the neighbors are preserved.  This significantly deviates from the
Node2Vec random walk model.  As will be shown in
Section~\ref{sec:results}, node classification accuracy suffers
drastically because of the problematic trim idea.

\item \emph{Poor efficiency}: Even with the much simplified graph
after trimming, the efficiency of Spark-Node2Vec is still poor.  There
are two main causes.  First, Spark RDDs are read-only; Recording
random walk steps in every iteration will result in a copy-on-write
and the creation of a new RDD.  Second, the computation performs
frequent join transformations.  However, a join often involves a
shuffle operation, which prepares data by sorting or hash
partitioning.  Unfortunately, shuffle often needs to spill the
intermediate data to disks, incurring significant disk I/O overhead.
As a result, Spark-Node2Vec has a difficult time processing even
mid-sized graphs, as will be shown in Section~\ref{sec:results}.


\end{list}

\section{Fast-Node2Vec}
\label{sec:graphlite}


We present Fast-Node2Vec in this section.  We begin by describing our
choice of graph computation framework in
Section~\ref{subsec:graphlite}.  Then, we describe the main components
of the Fast-Node2Vec algorithm in Section~\ref{subsec:fast-node2vec}.
After that, we analyze the computation to understand its efficiency
bottleneck in Section~\ref{subsec:analysis} and propose a number of
techniques to improve Fast-Node2Vec in Section~\ref{subsec:opt}.

\subsection{GraphLite: a Pregel-Like Distributed\\Graph Computation Framework}
\label{subsec:graphlite}

\begin{figure}
\centering{$
  \begin{array}{c}
  \hspace{-0.2in}
  \includegraphics[height=3.3cm]{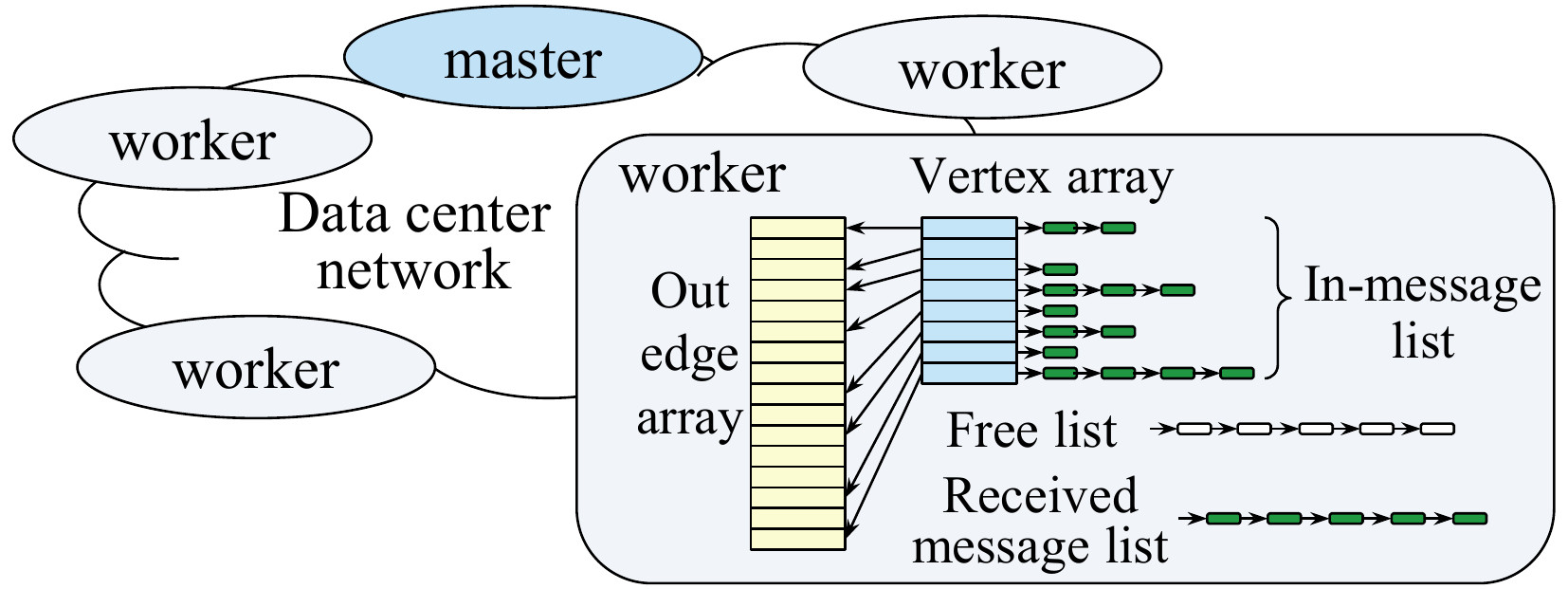}
  \end{array}
  \vspace{-0.1in}
$}
  \caption{GraphLite system architecture.}
  \label{fig:graphlite}
  \vspace{-0.1in}
\end{figure}

Pregel~\cite{MalewiczABDHLC10} is a distributed in-memory graph
computation framework.  GraphLite~\cite{NiuC15} is a lightweight
open-source C/C++ implementation of Pregel.  We choose GraphLite to
run Node2Vec in order to avoid the costs of RDDs and shuffles in
Spark, as discussed in Section~\ref{subsec:spark-node2vec}.


In the Pregel model, a graph algorithm is implemented as a
vertex-centric \insn{compute} function.  The system consists of a
master machine node and a number of worker nodes connected through a
data center network, as depicted in Figure~\ref{fig:graphlite}.  The
graph is partitioned across the workers at the beginning of the
computation.  The system runs the graph computation in a series of
supersteps.  At the beginning of a superstep, the master broadcasts a
\emph{start} message to all workers.  Upon receiving this message,
each worker iterates through all the vertices and invoke the
\insn{compute} function for every vertex in its graph partition.  Then
workers send \emph{done} messages to the master.  The master starts
the next superstep only after all the workers are done with the
current superstep.  In this way, the computation follows the Bulk
Synchronous Parallel model: The master enforces a global barrier
between two supersteps; Within each superstep, \insn{compute} is run
in parallel across workers.  

The \insn{compute} function on a vertex is typically implemented 
with three main parts: (i) receiving incoming messages from the
previous superstep; (ii) computing and updating the state of the
vertex in light of the incoming messages; (iii) sending messages to
other vertices that will be delivered in the next superstep.

Figure~\ref{fig:graphlite} illustrates the internal data structures of
a worker.  There are an array of vertex states, an array of vertex
out-edges, and structures for managing messages.  In a superstep,
incoming messages to the vertices in the local partition are appended
to the global \emph{received message list}.  At the start of the next
superstep, the messages are then delivered into the per-vertex
\emph{in-message lists}, which are processed in the \insn{compute}
function invoked later in the superstep.

Compared to Spark, the Pregel model has the following advantages.
First, vertex states are updated in place, while Spark incurs RDD
copy-on-write cost and consumes much more memory space.  Second,
vertices communicate through messages that are managed in main memory,
while Spark's shuffle operation may incur significant file I/O
overhead.  

\Paragraph{GAS model and Node2Vec} Apart from the Pregel model,
another popular graph computation model is GAS proposed in
PowerGraph~\cite{GonzalezLGBG12}.  The GAS model divides the
\insn{compute} function into three functions: Gather, Apply, and
Scatter.  While GAS can efficiently support many graph computation
tasks, such as PageRank, we find that GAS may not be suitable for
Node2Vec because the incoming messages in Node2Vec cannot be
aggregated and there are no broadcasting outgoing messages.
Consequently, the GAS approach does not introduce additional benefits
for Node2Vec.


\subsection{Fast-Node2Vec on GraphLite}
\label{subsec:fast-node2vec}


We aim to run Node2Vec efficiently on graphs with billions of vertices
and edges using a mid-sized cluster of machines.  Such computing
resources are often available to researchers in academia.  In our
experiments, we use a cluster of 12 machines, each with 128GB of
memory and 40 cores.  Since Pregel performs distributed computation in
memory, we aim to restrict the total memory consumption in Node2Vec to
the aggregate memory size in the cluster (i.e., 1.5TB).  

In the original Node2Vec~\cite{GroverL16}, all transition
probabilities are pre-computed before simulating the random walks.
This approach is followed by both the single-machine reference
implementations and Spark-Node2Vec.  However, this approach consumes a
large amount of memory.
Let us consider an undirected graph for simplicity.  Directed graphs
can be analyzed similarly.  Let the degree of a vertex $V_i$ be $d_i$.
As shown in Figure~\ref{fig:node2vec-random-walk}, there is an
$\alpha_{pq}(u,V_i,x)$ for each pair of $u$ and $x$, which are $V_i$'s
neighbors.  Thus, the number of $\alpha$'s at $V_i$ is equal to
$d_i^2$.  Alias sampling~\cite{Vose91} requires 8-byte space per
probability.  Therefore, the space for all the transition
probabilities can be computed as follows:
\vspace{-3mm} \begin{equation}
Memory_{TransProb} = 8\sum d_i^2 \geq 8\frac{(\sum d_i)^2}{n} = 8nd^2
\vspace{-1mm} \end{equation}
where $n$ is the number of vertices and $d$ is the average vertex
degree.  Here, we estimate the total amount of space required to store
all the transition probabilities of a social network.  Real-world
social networks (e.g., Twitter, Facebook, WeChat) often have over a
billion vertices.  The number of friends of an average user is on the
order of 100 to 1000.  Therefore, if $n=1G$ and $d=100$, it requires a
space of 80~TB.  If $n=1G$ and $d=1000$, it takes 8~PB to store all
transition probabilities.  The space required is clearly much larger
than the aggregate memory space available in a mid-sized cluster.

\begin{algorithm}[t]
    \caption{Fast-Node2Vec (FN-Base)}
    \label{alg:fn-base}
    \begin{algorithmic}[1]
    \Procedure{compute}{$msgIterator$, $walkLength$}
    \State $s= superstep$
    \If{$s == 0$}
        \State sample 1st step $x$ based on static edge weights
        \State save $x$ as $step[0]$ in vertex value 
        \State send $(NEIG, this.vid, neighbor\ vids)$ to vertex $x$
    \ElsIf{$s < walkLength$}
        \For{each $msg$ in $msgIterator$}
            \State /* message: (type, starting vid, value) */
            \If{$msg.type == STEP$ /* a step message */}
                \State save $msg.value$ as $step[s]$ in vertex value
            \Else{ /* a neighbor message */}
                \State group messages by the starting vid into $G$
            \EndIf
        \EndFor
        \For{each $group$ in $G$}
            \State compute transition probabilities
            \State sample the next step $x$
            \State $src$ = starting vid
            \State send $(STEP, src, x)$ to the vertex $src$
            \If{$s + 1 < walkLength$}
            \State send $(NEIG, src, neighbor\ vids)$ to vertex $x$
            \EndIf
        \EndFor
    \Else{ /* done with the current walk */}
        \State vote to halt
    \EndIf
    \EndProcedure
    \end{algorithmic}
\end{algorithm}

Given this observation, we propose to compute the transition
probability on demand during the biased random walk.
Algorithm~\ref{alg:fn-base} lists the pseudo-code for
Fast-Node2Vec on GraphLite.  There are several interesting design
choices in this algorithm.  First, the algorithm simulates $n$ random
walks, one per starting vertex, where $n$ is the number of vertices in
the graph.  At superstep $s$, it computes step $s$ for all $n$ random
walks.  Second, there are two types of messages: $STEP$ and $NEIG$.
The $STEP$ message reports a sampled step in a random walk.  The
$NEIG$ message sends the neighbors of the current vertex to the
next-step vertex $x$ in the random walk.  All messages are labelled
with the associated starting vertex ID (Line 9).  Third, the sampled
walk steps are stored in the value of the starting vertex (Line 5 and
11).  This is achieved by sending the $STEP$ message with the sampled
step as the value field to the starting vertex (Line 20).  Fourth, if
a random walk moves from $v$ to $x$, then the \insn{compute} at $v$
will send $v$'s neighbor vertex IDs to $x$ (Line 22).  In this way,
$x$ can easily figure out the shared neighbors between $x$ and $v$ for
the transition probability computation (Line 17).  Finally, we note
that multiple random walks may arrive at the same vertex in some step.
The algorithm deals with this complexity by grouping messages based on
the starting vertex IDs (Line 13 and 16).


We now explain how to compute transition probabilities.  A vertex can
access its outgoing edges in the out-edge array in GraphLite to get
its neighbors and edge weights.  Suppose step $s-1$ of the random walk
is at $u$ and the current step $s$ is at $v$. We would like to compute
the transition probabilities at $v$.  According to
Figure~\ref{fig:node2vec-random-walk}, we need to figure out the
distance $dist(u,x)$ between $u$ and every neighbor $x$ of $v$.  In
Algorithm~\ref{alg:fn-base}, in superstep $s-1$, \insn{compute} at
$u$ has already sent $u$'s neighbors to $v$ (Line 22).  Therefore, in
superstep $s$, \insn{compute} at $v$ will receive $u$'s neighbors in
incoming messages (Line 12-14).  The three cases of distances in
Figure~\ref{fig:node2vec-random-walk} means that $x$ is $u$, $x$ is a
common neighbor of $u$ and $v$, and all other cases.  It is easy to
find out if $x$ is a common neighbor of $u$ and $v$ by using a hash
set.  In this way, we obtain the unnormalized transition
probabilities, then use them to perform the biased sample at $v$ (Line
18).

%


\subsection{Analysis of Fast-Node2Vec Computation}
\label{subsec:analysis}

\begin{figure}[t]
\centering{$
  \begin{array}{c}
    \hspace{-0.1in}
  \includegraphics[width=8.8cm]{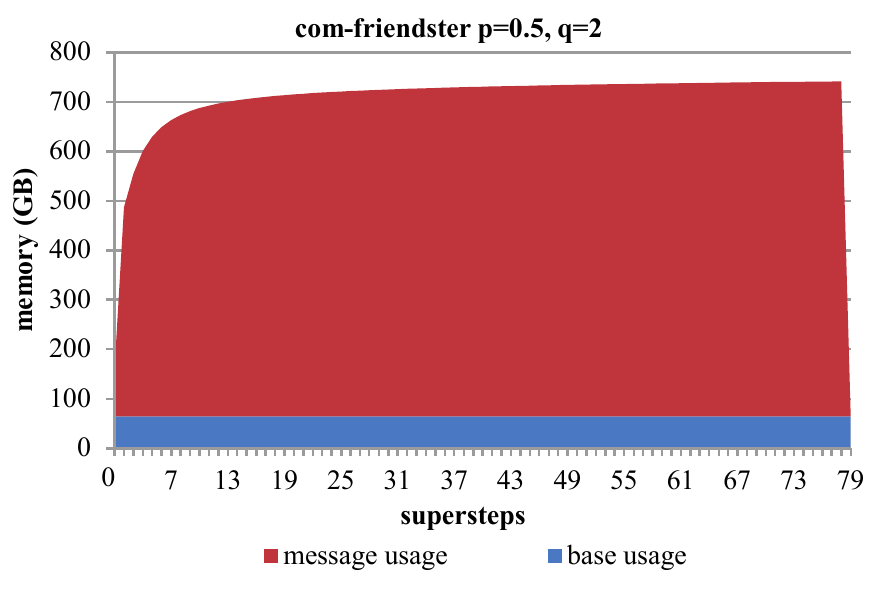}
  \end{array}
  \vspace{-0.2in}
$}
  \caption{The memory space consumed increases then flattens.
(FN-Base, com-Friendster data set)}
  \label{fig:mem-friendster}
  \vspace{-0.1in}
\end{figure}

\begin{figure}[t]
\centering{$
  \begin{array}{c}
    \hspace{-0.1in}
    \includegraphics[width=8.8cm]{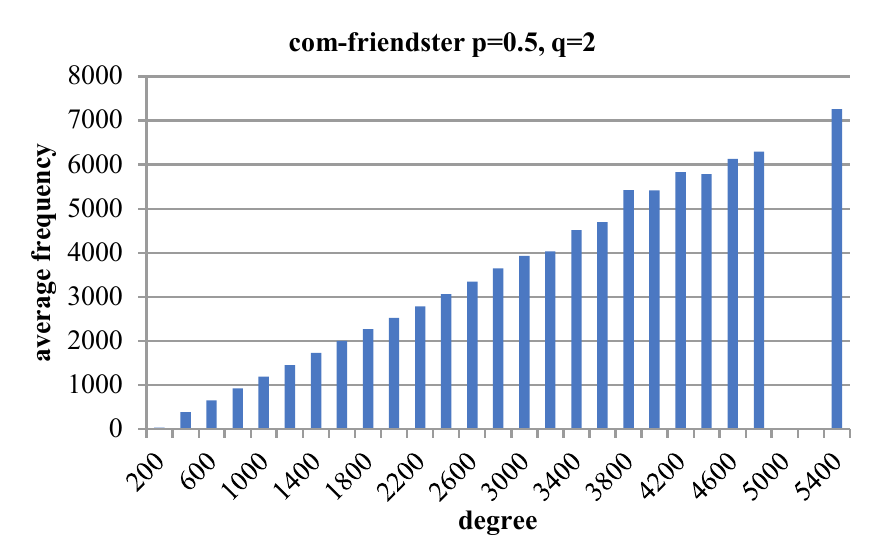}
  \end{array}
  \vspace{-0.2in}
$}
    \caption{The average frequencies of vertices that
appear in the random walks vs. the degrees of the vertices. 
(FN-Base, com-Friendster data set)}
  \label{fig:freq-friendster}
  \vspace{-0.1in}
\end{figure}

Figure~\ref{fig:mem-friendster} shows the change in the memory space
consumed during the Fast-Node2Vec computation on com-Friendster, the
largest real-world graph in our evaluation (c.f.
Section~\ref{subsec:setup}).  We see that the amount of memory
consumed quickly increases, and then flattens after about 10
supersteps.  To understand this behavior, we consider the memory
consumed during graph computation.  There are mainly three parts that
consume significant amount of memory: (i) the graph topology,
including vertices and edges; (ii) the vertex values and edge weights;
(iii) the $STEP$ and $NEIG$ messages.  In the Node2Vec computation,
the graph topology is unchanged and the memory space allocated for the
vertex values and the edge weights keeps the same.  Therefore, part
(i) and part (ii), shown as base usage in
Figure~\ref{fig:mem-friendster}, do not contribute to the increase in
the observed memory consumption.  On the other hand, the memory
consumed by messages grow significantly, as shown in
Figure~\ref{fig:mem-friendster}.  The number of $STEP$ messages is
equal to $n$, the number of vertices, in Fast-Node2Vec.  The size of a
$STEP$ message is always the same.  Therefore, the memory consumed by
all $STEP$ messages is the same in the entire computation.  In
contrast, the $NEIG$ message sizes change significantly.  

We analyze the relationship of the vertex degrees and the frequency of
vertices to appear in the resulting random walks, as shown in
Figure~\ref{fig:freq-friendster}.  The X-axis shows equi-width buckets
of degrees.  For example, the bucket 600 contains all vertices with
degrees between 400 and 600.  The height of the bar represents the
average times for a vertex in this bucket to be sampled in the
Node2Vec random walks.  From Figure~\ref{fig:freq-friendster}, we see
that the higher the vertex degree, the more frequently that the vertex
is sampled in the random walks.    

We can explain this phenomenon as follows.  For a vertex $v$,
if a random walk arrives at one of $v$'s neighbors, there is a chance
for the random walk to move to $v$.   The higher the degree of $v$,
the more neighbors that $v$ has, the more likely that the previous
vertex in the random walk is a neighbor of $v$.  This effect means
that random walks will tend to visit large-degree vertices.   We know
that the higher the degree of $v$, the larger size $v$'s $NEIG$
messages take.  At the beginning, there is a walk at every vertex.
Gradually, more and more large-degree vertices appear in the walk.
Therefore, the memory consumed by $NEIG$ messages increases.  When
this behavior becomes stable, the memory consumed becomes flat.

While Figure~\ref{fig:mem-friendster} and
Figure~\ref{fig:freq-friendster} show the results for one set of
Node2Vec $(p,q)$ parameters, we see similar behaviors across different
parameter settings.

%

\subsection{Optimization Techniques}
\label{subsec:opt}

We have seen that the neighbor messages can consume a large amount of
memory.  This is especially the case for power-law graphs where a
small number of vertices have very large degrees.  We call such
vertices \emph{popular} vertices.  The memory consumption effectively
determines the graph sizes that can be supported in a mid-sized
cluster.  In this sub-section, we propose a number of techniques to
reduce the message sizes for saving memory and improve the efficiency
of Fast-Node2Vec.

\vspace{2mm}
\Paragraph{FN-Local: Exploiting local graph partition}  Consider an
$NEIG$ message sent from $u$ to $v$.  The purpose of this message is
to notify $v$ of $u$'s neighbors.  This is only necessary if $u$ and
$v$ are in separate workers.  If both $u$ and $v$ are in the same
worker, $v$ is able to directly read $u$'s neighbors from the out-edge
array.  We call an $NEIG$ message between vertices in the same worker
a \emph{local} $NEIG$ message.  We extend the GraphLite framework with
an API that allows a vertex to visit another vertex's information in
the same worker.  Then FN-Local uses this API to reduce all the local
$NEIG$ messages.

\vspace{2mm}
\Paragraph{FN-Switch: Switching the destination of $NEIG$ messages
from popular vertices to unpopular vertices}  The NEIG message size is
proportional to the number of neighbors of a vertex.  Consider an
$NEIG$ message from $u$ to $v$, where $u$ is a popular vertex and $v$
is an unpopular vertex.  The purpose of this message is to facilitate
the computation of common neighbors between $u$ and $v$.  However,
since a popular vertex has a large number of neighbors, the message
size from $u$ to $v$ is very large.

An interesting observation is that the common neighbors between $u$
and $v$ can also be computed at $u$ if $v$ sends all its neighbors to
$u$.  In addition, after $u$ knows all $v$'s neighbors, it can perform
the full computation of transition probabilities and random walk
simulation on behalf of $v$.   In other words, it is possible to
switch the destination of the $NEIG$ message, asking $v$ to send its
neighbors to $u$.  Since $v$ has much fewer neighbors than $u$, this
idea may significantly reduce the message size.

While this destination switching idea seems promising at first sight,
back-of-envelope computation shows that it could significantly
increase the overall run time.  There is a significant problem to
implement this idea.  There are two messages for computing the random
walk step: $u$ must notify $v$ that it wants $v$ to send back an
$NEIG$ message;  Then $v$ sends the $NEIG$ message to $u$.  This means
that the random walk step needs \emph{two} supersteps.  This breaks
Fast-Node2Vec's behavior that it computes the same step of all random
walks in every superstep.  Consider a random walk that consists of
alternating popular and unpopular vertices.  Half of the moves are
from a popular vertex to an unpopular vertex.  Every such move takes
one extra superstep.  Thus, the entire random walk will take 50\% more
supersteps to complete, incurring significant time overhead. 

\vspace{2mm}
\Paragraph{FN-Cache: Caching neighbors of remote popular vertices} If
a popular vertex $u$ sends its neighbors to a vertex $v$ in a remote
worker $W$, $v$ can cache $u$'s neighbors in a global data structure
at worker $W$ so that later computation at any vertices in worker $W$
can directly access this information without $u$ re-sending costly
$NEIG$ messages.  

To implement this idea, we extend GraphLite to expose an API for
looking up the worker ID of a vertex.  A popular vertex $u$ will
remember in an $WorkerSent$ set to which remote workers it has sent
$NEIG$ messages.  Before sending an $NEIG$ message, $u$ checks to see
if the destination vertex $v$'s worker is in the $WorkerSent$ set.  If
no, $u$ sends a normal $NEIG$ message and records $v$'s worker ID in
the $WorkerSent$ set.  If yes, $u$ sends a special $NEIG$ message with
a special (otherwise unused) value to notify $v$ to look up $u$'s
neighbors locally.  In this way, this technique can significantly
improve the efficiency of Fast-Node2Vec as will be shown in
Section~\ref{sec:results}.

\vspace{2mm}
\Paragraph{FN-Multi: Simulating random walks using multiple rounds}
After the above optimizations, it is possible that the memory space
required by Fast-Node2Vec is still too large to fit into the aggregate
memory size in the machine cluster.  We would like to gracefully
handle such situations.  We observe that the random walks starting
from different vertices are independent of each other.  Therefore, it
is not necessary to run all $n$ random walks at the same time.
Instead, we can simulate the random walks in $k$ rounds. In each
round, we simulate $n/k$ random walks.  
This technique will reduce the memory space for managing messages and
for recording random walks by about a factor of $k$ times.  Note that
we cannot remove any vertex or edge from the graph to reduce memory
space for vertices and edges.  This is because every vertex and every
edge may still be visited in the subset of random walks.

\vspace{2mm}
\Paragraph{FN-Approx: Approximately computing the random walk steps at
popular vertices}  The cost of computing the transition probabilities
and simulating a random walk step at vertex $v$ is $O(d_v)$, where
$d_v$ is $v$'s degree.   Therefore, popular vertices take much longer
time for this computation.  We would like to reduce the computation
cost at popular vertices.

Consider an $NEIG$ message from an unpopular vertex $u$ to a popular
vertex $v$.  We can derive the upper and lower bounds for an
individual transition probability at $v$.  Suppose $\frac{1}{p} \leq 1
\leq \frac{1}{q}$.  Let $d_u$ be $u$'s degree.  Then the bounds for
the transition probability from $v$ to a neighbor $x$ that is not $u$
are as follows: 
\vspace{-0.02in} \begin{equation}
LowerBound =
\frac{Weight_{min}}{(\tfrac{1}{p}+\tfrac{d_v-1}{q})Weight_{max}}\\
\end{equation}
\vspace{-0.02in}
\begin{equation}
UpperBound = \frac{\tfrac{Weight_{max}}{q}}{( \tfrac{1}{p}+ d_u+
\tfrac{d_v-d_u-1}{q} )Weight_{min}}\\
\vspace{-0.02in} \end{equation}
The numerator is the unnormalized transition probability.  Its minimal
value is $Weight_{min}$, which is the product of the minimal
$\alpha=1$ (when $x$ is not $u$) and the minimal edge weight.   Its
maximal value is $\tfrac{Weight_{max}}{q}$, which is the product of
the maximal $\alpha$ and the maximal edge weight.  For the
denominator, the minimal value is achieved when all of $u$'s $d_u$
neighbors are also $v$'s neighbors.  The maximal value is achieved
when $u$ and $v$ do not have common neighbors.  Similarly, we can
obtain the upper and lower bounds for other value combinations of $p$
and $q$.

Note that $d_u << d_v$.  In many real-world scenarios, the difference
between $Weight_{min}$ and $Weight_{max}$ is small.  (For example, in
a great many cases, edge weights are 1 for all edges).  Therefore, the
above lower bound is close to $q/d_v$, and the above upper bound is
close to $1/d_v$.  As $d_v$ is very large, the difference between the
lower and the upper bounds can be very small.  

Base on this observation, we propose FN-Approx, an approximate
Fast-Node2Vec algorithm.   At a popular vertex $v$, if the last step
$u$ is an unpopular vertex, FN-Approx computes the difference between
the upper and the lower bounds.  If the difference is below a
pre-defined threshold (e.g., 1e-3), FN-Approx will simply sample the
step based on the static edge weights without considering the 2nd
order effect.  In this way, FN-Approx can significantly reduce the
time overhead for computing transition probabilities at popular
vertices.  We study the impact of this approximation on the efficiency
and the accuracy of the solution in Section~\ref{sec:results}.

%
%
%
%

\section{Evaluation}
\label{sec:results}


In this section, we empirically evaluate the efficiency of our
proposed Fast-Node2Vec algorithms.  We apply the algorithms to a
number of large-scale real-world and synthetic graphs. 

\subsection{Experimental Setup}
\label{subsec:setup}

\Paragraph{Machine Configuration} We run all experiments on a cluster
of 12 machines, each of which is equipped with two Intel(R) Xeon(R)
E5-2650 v3 CPU @ 2.30GHz (10 cores, 2 threads/core) and 128GB DRAM.
The machine runs stock Ubuntu 16.04 with Linux Kernel version
4.4.0-112-generic. The machine nodes are connected through 10~Gbps
Ethernet.  The measured network bandwidth is between 9.4 and 9.6~Gbps.
We compile C-Node2Vec, GraphLite, and the Fast-Node2Vec algorithms
using g++ version 5.4.0 with optimization level -O3. For
Spark-Node2Vec, we run Spark version 2.2.0 with Java version 1.8 and
Scala version 2.11.  We deploy a cluster of 11 Spark workers and 1
Spark master and set the driver memory and executor memory size as
100~GB for Spark-Node2Vec evaluation to utilize almost all memory
available on the machine nodes.

\Paragraph{Data Sets} Table~\ref{tab:datasets} summarizes the
real-world and synthetic data sets used in the evaluation.  The
real-world graphs are described in the following:
\begin{list}{\labelitemi}{\setlength{\leftmargin}{5mm}\setlength{\itemindent}{0mm}\setlength{\topsep}{0.5mm}\setlength{\itemsep}{1mm}\setlength{\parsep}{0.5mm}}

\item BlogCatalog~\cite{zafarani2009social}: This is a social network
between authors on the BlogCatalog site.  Vertices are labelled with
topic categories provided by authors.  This is the same data set used
in the Node2Vec paper~\cite{GroverL16}.  We use This data set to
evaluate not only the efficiency the algorithms, but also the accuracy
of the resulting vector representations for the node classification
task.  


\item com-LiveJournal~\cite{YangL12}: This is the friendship social
network of the LiveJournal blogging website.  There are 4 million
vertices and 34.7 million edges in the graph.

\item com-Orkut~\cite{YangL12}: This is a social network of the Orkut
site.  Compared to com-LiveJournal, this graph contains slightly
smaller number of vertices (3.1 million) but much larger number of
edges (117.2 million).  The average vertex degree of com-Orkut is 4.3
times as large as that of com-LiveJournal.

\item com-Friendster~\cite{YangL12}: This is a network of social
relationships of the users on the Friendster site.  Containing 1.8
billion edges, com-Friendster is the largest real-world data set in
the evaluation.

\end{list}

\begin{table}[t]
    \vspace{-2mm}
    \centering
    \caption{Graphs used in the evaluation.}
    \label{tab:datasets}
    \small
    \begin{tabular}{|l|c|c|c|c|}  \hline
    Graph            & $\|V\|$           & $\|E\|$& Max Degree\\ \hline\hline
    BlogCatalog      & 10.3K             & 334.0K & 3,854     \\
    com-LiveJournal  & 4.0M              & 34.7M  & 14,815    \\
    com-Orkut        & 3.1M              & 117.2M & 58,999    \\
    com-Friendster   & 65.6M             & 1.8G   & 8,447     \\\hline
    ER-20            & 1.0M ($2^{20}$)   & 10.5M  & 29        \\
    ER-22            & 4.2M ($2^{22}$)   & 41.9M  & 31        \\
    ER-24            & 16.8M ($2^{24}$)  & 167.8M & 31        \\
    ER-26            & 67.1M ($2^{26}$)  & 671.1M & 32        \\
    ER-28            & 268.4M ($2^{28}$) & 2.7G   & 33        \\
    ER-30            & 1.1G ($2^{30}$)   & 10.7G  & 35        \\ \hline
    WeC-20           & 1.0M ($2^{20}$)   & 104.8M & 1,053     \\
    WeC-22           & 4.2M ($2^{22}$)   & 419.4M & 1,745     \\
    WeC-24           & 16.8M ($2^{24}$)  & 167.8M & 2,316     \\
    WeC-26           & 67.1M ($2^{26}$)  & 6.7G   & 2,771     \\ \hline
    Skew-1           & 4.2M ($2^{22}$)   & 419.4M & 159       \\
    Skew-2           & 4.2M ($2^{22}$)   & 419.4M & 2,098     \\
    Skew-3           & 4.2M ($2^{22}$)   & 419.4M & 8,420     \\
    Skew-4           & 4.2M ($2^{22}$)   & 419.2M & 23,594    \\
    Skew-5           & 4.2M ($2^{22}$)   & 418.3M & 36,914    \\\hline
    \end{tabular}
\end{table}

\begin{figure*}
\centering{$
  \begin{array}{c}
  \hspace{-0.1in}
  \includegraphics[height=5.1cm]{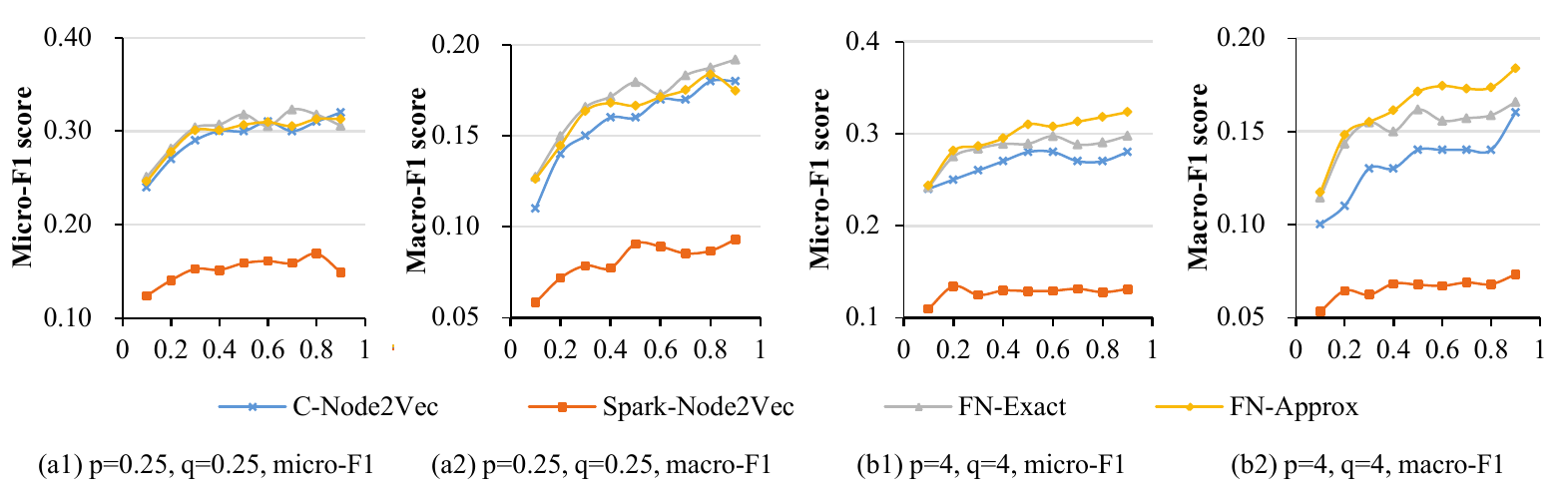}
  \end{array}
  \vspace{-0.15in}
$}
  \caption{The accuracy of the node classification task on
the BlogCatalog graph. (X-axis denotes the fraction of labelled data
used as the training data set.)}
  \label{fig:multi-label-classification}
\end{figure*}

\noindent In addition to the real-world graphs, we generate synthetic
graphs that follow the RMAT~\cite{ChakrabartiZF04} model.  In the RMAT
model, a graph with $2^K$ vertices is generated using four parameters
($a$, $b$, $c$, $d$), where $a+b+c+d=1$.  The $2^K\times2^K$ adjacency
matrix is conceptually divided into four $2^{K-1}\times2^{K-1}$
sub-matrices.  An edge is randomly generated in the top-left,
top-right, bottom-left, and bottom-right sub-matrices with probability
$a$, $b$, $c$, and $d$, respectively.  This process is recursively
applied to every sub-matrix.  That is, a $2^{i}\times2^{i}$ matrix is
conceptually divided into four $2^{i-1}\times2^{i-1}$ matrices, where
$i=K,K-1,\cdots,2$.  Given that an edge is to be generated in a
$2^{i}\times2^{i}$ matrix, the probability that this edge belongs to
one of its four $2^{i-1}\times2^{i-1}$ sub-matrices follows the ($a$,
$b$, $c$, $d$) distribution.  We use a graph generation tool called
TrillionG\footnote{\small\url{https://github.com/chan150/TrillionG}}~\cite{Park017}
to generate large-scale RMAT graphs.  We vary ($a$, $b$, $c$, $d$) to
generate three sets of graphs with different characteristics:  
\begin{list}{\labelitemi}{\setlength{\leftmargin}{5mm}\setlength{\itemindent}{0mm}\setlength{\topsep}{0.5mm}\setlength{\itemsep}{1mm}\setlength{\parsep}{0.5mm}}

\item ER-K graphs: We set ($a$, $b$, $c$, $d$) to (0.25, 0.25, 0.25,
0.25) to generate Erdos-Renyi (ER) graphs with $2^K$ vertices and an
average degree of 10.  Note that the edges in this graph are uniformly
distributed.  There is no skew in the vertex degree distribution.  We
use this graph because the Node2Vec paper~\cite{GroverL16} reports run
times for ER graphs with up to 1 million vertices and 10 million
edges.  We would like to compare our solution with the original
Node2Vec, and show that our solution can scale up to 1 billion
vertices and 10 billion edges.  Therefore, we vary $K$ from 20 to 30
to generate graphs with 1 million to 1.1 billion vertices.

\item WeC-K graphs: We model a WeChat-like social network, where the
upper bound of the number of friends per user is 5,000 and the average
number of friends is about 100.  We generate a set of WeC-K graphs
with $2^K$ vertices and $100\times2^K$ edges.  Without loss of
generality, suppose $c+d \geq a+b$.  Then the vertex with the largest
degree is the last vertex with high probability.  We can compute the
parameters to ensure the expected degree of the last vertex is 5000.
While the resulting parameters vary slightly for different $K$, we
choose (0.18, 0.25, 0.25, 0.32) as the representative parameters to
generate all WeC-K graphs so that the graphs are comparable to the
Skew-S graphs.

%

\item Skew-S graphs: We generate a set of graphs with $2^{22}$
vertices and an average degree of 100, while varying the skewness of
the data.  We set the parameters $(a,b,c,d)$ so that the number of
edges in the bottom-right part of the matrix is about $S$ times as
many as that in the top-left part of the matrix, i.e. $d = Sa$.  In
addition, we set $b = c = 0.25$.  When $S=1$, there is no skew.  In
general, when $S>1$, RMAT generates graphs with power-law
characteristics. The higher the $S$, the more skew the vertex degrees
are.  We vary $S$ from 1 to 5.  Note that WeC-22 is Skew-1.78
($0.32/0.18=1.78$).

\end{list}    


\Paragraph{Algorithms to Compare} We evaluate the following solutions
in the experiments:
\begin{list}{\labelitemi}{\setlength{\leftmargin}{5mm}\setlength{\itemindent}{0mm}\setlength{\topsep}{0.5mm}\setlength{\itemsep}{1mm}\setlength{\parsep}{0.5mm}}

\item C-Node2Vec: This is the single-machine C++ reference
implementation from the Node2Vec project web page.  We use this
implementation for two purposes: (i) comparing accuracy of applying
the various solutions to the node classification task; and (ii)
evaluating the scalability using the ER-K graphs.

\item Spark-Node2Vec: This is the Node2Vec implementation on Spark.
It preserves up to 30 edges per vertex, and computes the transition
probabilities before running the biased random walk.


\item FN-Base: This is Algorithm~\ref{alg:fn-base}, the baseline
Fast-Node2Vec algorithm, which computes transition probabilities on
demand and runs on GraphLite, a Pregel-Like Graph Computation
Framework (cf. Section~\ref{subsec:fast-node2vec}).

\item FN-Local: FN-Local improves FN-Base by allowing vertices to
visit edge information of other vertices in the local graph
partitions (cf. Section~\ref{subsec:opt}).  In this way,
it reduces the $NEIG$ messages for local vertices.

\item FN-Switch: FN-Switch switches the destination of $NEIG$ messages
from popular to unpopular vertices (cf. Section~\ref{subsec:opt}).
However, this may incur more rounds of messages and thus more
supersteps.

\item FN-Cache: FN-Cache improves FN-Local by caching neighbors of
the most popular vertices in order to reduce the amount of messages
for popular vertices (cf. Section~\ref{subsec:opt}).


\item FN-Approx and FN-Exact: FN-Approx is the approximate algorithm
to reduce the overhead for computing transition probability at popular
vertices.  It is otherwise the same as FN-Cache (cf.
Section~\ref{subsec:opt}).  As the other FN algorithms all give exact
results, we use FN-Exact to represent them for the purpose of
comparing accuracy of applying the solutions to the node
classification task.

\end{list}

\Paragraph{Measurements} We compare both the result quality and the
efficiency of the above solutions in our experiments.  In the
efficiency comparison, we will focus only on the run time of
performing the Node2Vec random walks.  For each combination of
solutions, graphs, and Node2Vec $(p,q)$ parameter settings, we compute
80-step biased random walks for all vertices in the graph.

\begin{figure}[t]
\centering{$
  \begin{array}{c}
  \hspace{-3mm}
  \includegraphics[height=5.80cm]{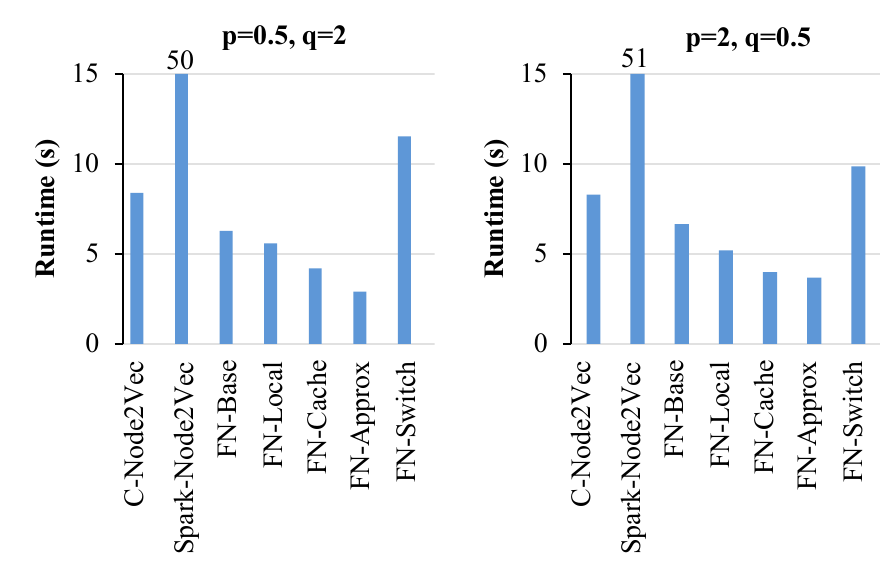}
  \vspace{-0.10in}\\
  \mbox{(a) BlogCatalog}
  \vspace{0.05in}\\
  \hspace{-3mm}
  \includegraphics[height=5.80cm]{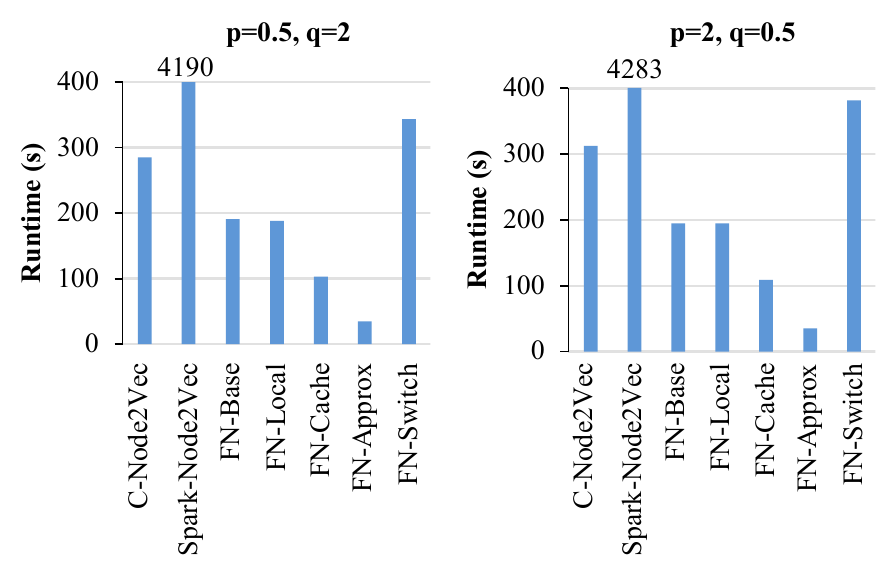}
  \vspace{-0.10in}\\
  \mbox{(b) com-LiveJournal}
  \vspace{0.05in}\\
  \hspace{-3mm}
  \includegraphics[height=5.80cm]{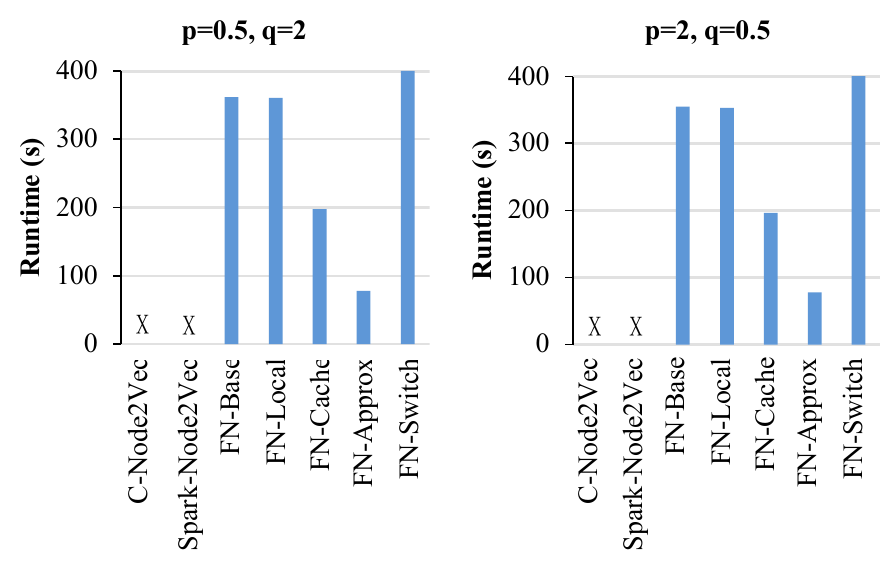}
  \vspace{-0.10in}\\
  \mbox{(c) com-Orkut}\\
  \end{array}
  \vspace{-0.1in}
$}
  \caption{Comparing efficiency of the solutions using real-world
graphs (BlogCatalog, com-LiveJournal, and com-Orkut) (``x'' means that
the program runs out of memory and is killed by the OS.)}.
  \label{fig:real}
  \vspace{-0.1in}
\end{figure}

\begin{figure}[t]
\centering{$
  \begin{array}{c}
  \hspace{-0.15in}
  \includegraphics[width=8.0cm]{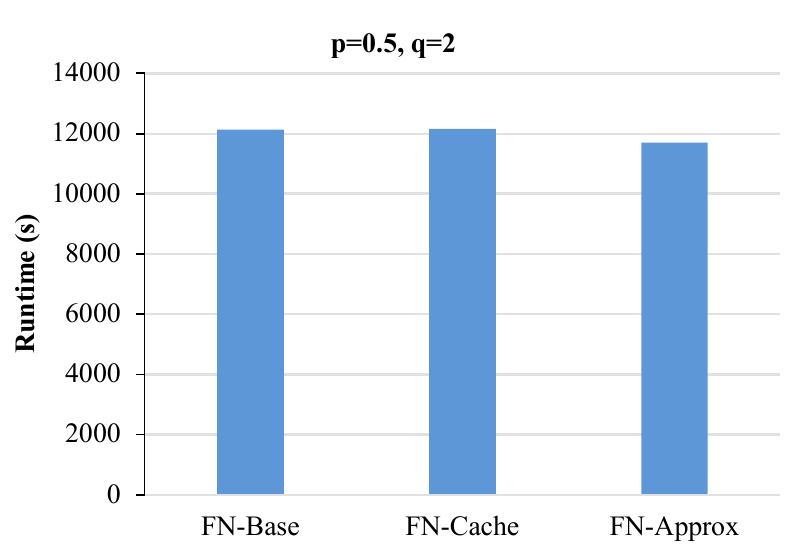}
  \end{array}
  \vspace{-0.15in}
$}
  \caption{com-Friendster.}
  \label{fig:friendster}
  \vspace{-0.1in}
\end{figure}

\subsection{Accuracy of Node Classification}
\label{subsec:accuracy}

The first question that we would like to answer is what is the impact
of the quality of the generated random walks on the accuracy of the
node classification task.  We use the BlogCatalog data set, in which
vertex labels are available, for this purpose.
Figure~\ref{fig:multi-label-classification} compares the accuracy of
the node classification task using the vector representations
generated by C-Node2Vec, Spark-Node2Vec, FN-Exact, and FN-approx.
From left to right, the figure shows the micro-F1 and macro-F1 scores
with two sets of Node2Vec $(p, q)$ parameters.  The higher the scores,
the better the solution.

From Figure~\ref{fig:multi-label-classification}, we see that: 
\begin{list}{(\arabic{enumi})}{\setlength{\leftmargin}{5mm}\setlength{\itemindent}{0mm}\setlength{\topsep}{0.5mm}\setlength{\itemsep}{1mm}\setlength{\parsep}{0.5mm}}

\setcounter{enumi}{0}

\addtocounter{enumi}{1}
\item \emph{The accuracy of Spark-Node2Vec is dramatically worse than
the other solutions}.  This is because Spark-Node2Vec preserves only
up to 30 edges for every vertex in order to save memory space,
significantly restricting the random walk destinations and thus
altering the behavior of Node2Vec random walks.  

\addtocounter{enumi}{1}
\item \emph{FN-exact achieves similar accuracy as C-Node2Vec}.
FN-exact (i.e., FN-Base, FN-Local, FN-Switch, FN-Cache) implements the
2nd-order biased random walk exactly as defined in
Node2Vec~\cite{GroverL16}.  Therefore, it has similar quality as the
reference implementation, C-Node2Vec.

\addtocounter{enumi}{1}
\item \emph{Interestingly, FN-approx, our proposed approximate
solution, achieves similar accuracy compared to FN-exact and
C-Node2Vec}.  This shows that the quality degradation caused by
the approximate computation on popular vertices is neglegible.
This approximation technique works in practice.

\end{list}

\subsection{Efficiency on Real-World Graphs}
\label{subsec:real}

Figures~\ref{fig:real}(a)-(c) compare the execution time of all seven
solutions: C-Node2Vec, Spark-Node2Vec, FN-Base, FN-Local, FN-Cache,
FN-Approx, and FN-Switch.  For each graph, we run experiments with
two sets of Node2Vec $(p,q)$ parameters.  From the figures, we can
see the following trends:
\begin{list}{(\arabic{enumi})}{\setlength{\leftmargin}{5mm}\setlength{\itemindent}{0mm}\setlength{\topsep}{0.5mm}\setlength{\itemsep}{1mm}\setlength{\parsep}{0.5mm}}

\setcounter{enumi}{0}

\addtocounter{enumi}{1}
\item \emph{C-Node2Vec, a single-machine solution, cannot support very
large graphs, such as com-Orkut}.  When the space required by the
graph and the Node2Vec algorithm is too large to fit into the memory
of a single machine, it is desirable to run distributed Node2Vec
solutions.

\addtocounter{enumi}{1}
\item \emph{Spark-Node2Vec is by far the slowest solution}.
Spark-Node2Vec tries to reduce memory space by restricting the number
of edges per vertex to 30.  As shown in Section~\ref{subsec:accuracy},
this trick drastically degrades the quality of the solution.  However,
even with this trick, Spark-Node2Vec still runs out of memory for
com-Orkut on our 12-node cluster.  Moreover, it suffers from the
inefficiency of read-only RDDs and the I/O intensive shuffle
operations. 

\addtocounter{enumi}{1}
\item \emph{FN-Base achieves 7.7-22x speedups over Spark-Node2Vec for
the cases that Spark-Node2Vec can support.} The improvements of
FN-Base are twofold. First, FN-Base employs a Pregel-like graph
computation platform that avoids the overhead of RDDs and disk I/Os in
Spark.  Second, FN-Base computes the transition probabilities on the
fly, thereby significantly reducing the memory required to store the
transition probabilities.  Suppose each probability requires 8-byte
space.  Then, the total amount of memory saved for storing the
transition probabilities is 3.0GB, 68.6GB, 731.9GB for BlogCatalog,
com-LiveJournal, and com-Orkut, respectively.


\addtocounter{enumi}{1}
\item \emph{FN-Local's execution time is quite similar to that of
FN-Base}.  While FN-Local reduces the NEIG messages between vertices
in the same graph partition, we find that the direct memory visits at
a vertex $u$ to retrieve the edge information of another vertex $v$
may incur expensive CPU cache misses, which is especially the case
when $v$'s degree is small.  As a result, the overall benefit is not
as large as we expected.

\addtocounter{enumi}{1}
\item \emph{FN-Cache and FN-Approx achieve 1.5--1.9x and 1.8--5.6x
speedups over FN-Base, respectively}.  FN-Cache employs caching for
the edge information of popular vertices.  In this way, it
significantly reduces the cost of NEIG messages.  FN-Approx performs
approximate computation (random sampling) on popular vertices when the
estimation errors are low.  This further reduces the computation
overhead.  Overall, \emph{compared to Spark-Node2Vec, FN-Cache and
FN-Approx achieve 11.9--40.8x and 13.8--122x speedups, respectively}.

\addtocounter{enumi}{1}
\item \emph{FN-Switch has the longest run time among the Fast-Node2Vec
solutions}.  The switch of an NEIG message requires an additional
message to be sent, thereby increasing the total number of supersteps
in the graph computation.  This effect significantly offsets the
potential benefit of switching the NEIG messages, resulting in poor
efficiency.

\end{list}

\noindent  From Figure~\ref{fig:multi-label-classification} and
Figure~\ref{fig:real}, we can conclude that Spark-Node2Vec has both
poor result quality and poor efficiency.  Moreover, among the
Fast-Node2Vec solutions, FN-Switch has poor efficiency and FN-Local
achieves similar execution time as FN-Base.  Therefore, we will not
consider Spark-Node2Vec, FN-Local, and FN-Switch any more in the rest
of the evaluation.

\Paragraph{The com-Friendster Graph} Figure~\ref{fig:friendster}
compares the execution time of FN-Base, FN-Cache, and FN-Approx for
the com-Friendster Graph.  com-Friendster is the largest real-world
graph with 1.8 billion edges.  The total amount of space required to
store all the transition probabilites is 11.6TB, much larger than the
total memory size ($\sim$1.5TB) of our 12-node cluster.  Therefore,
the pre-computation approach is not possible.  Computing the
transition probabilities on the fly is a must for larger graphs.
Taking this approach, FN-Base computes the full Node2Vec random walks
in about 3.3 hours.  This shows that FN-Base is capable of processing
large-scale real-world graphs in a reasonable amount of time using a
mid-sized cluster of machines.

However, FN-Base has already consumes the majority of memory for
com-Friendster.  Hence, FN-Cache has only limited amount of memory
available for caching edge information for popular vertices, which
does not show significant benefit.

\subsection{Scalability on ER-K Graphs}
\label{subsec:ER}

\begin{figure}[t]
\centering{$
  \begin{array}{c}
  \hspace{-0.15in}
  \includegraphics[width=8.9cm]{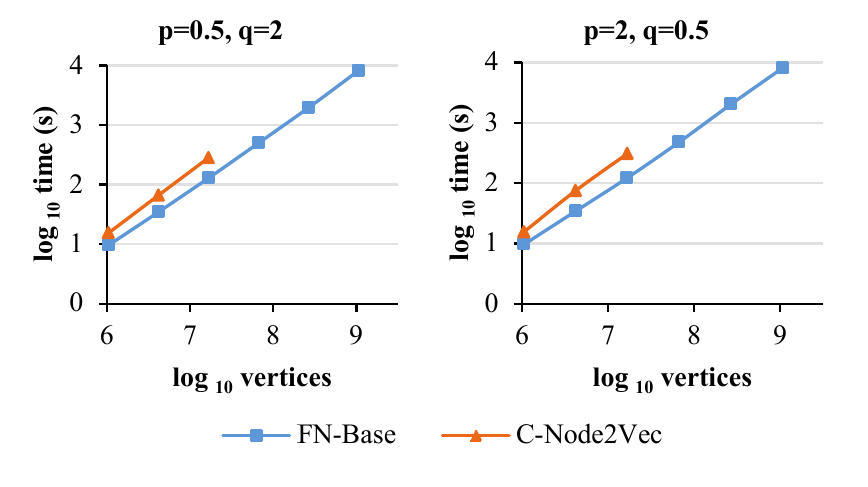}
  \end{array}
  \vspace{-0.2in}
$}
  \caption{Scalability of the Fast-Node2Vec solution using ER-K
graphs.}
  \label{fig:ER}
  \vspace{-0.1in}
\end{figure}

\begin{figure}[t]
\centering{$
  \begin{array}{c}
  \hspace{-5mm}
  \includegraphics[height=4.5cm]{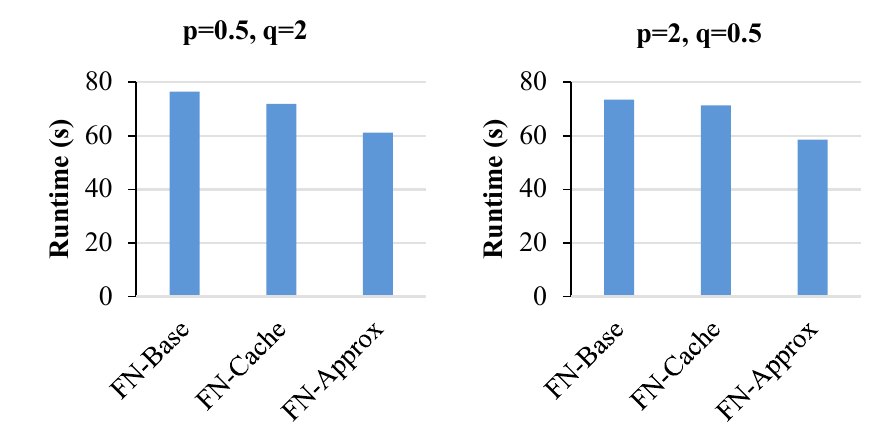}
  \vspace{-2mm}
  \\
  \mbox{(a) WeC-20}
  \\
  \hspace{-5mm}
  \includegraphics[height=4.5cm]{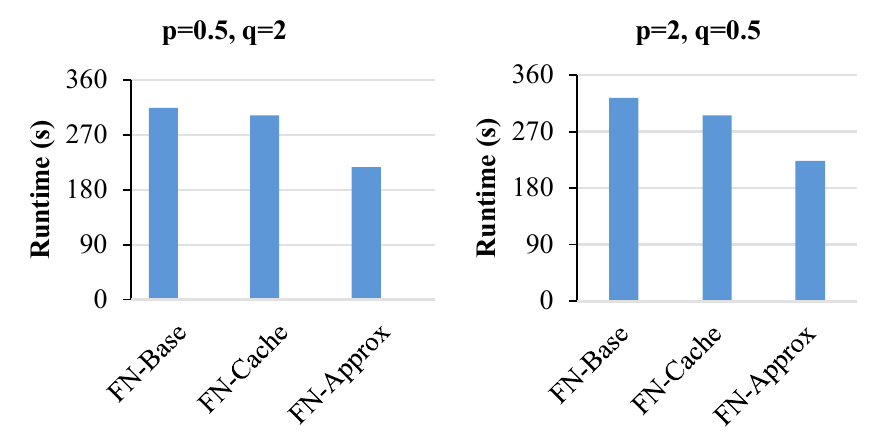}
  \vspace{-2mm}
  \\
  \mbox{(b) WeC-22}
  \\
  \hspace{-5mm}
  \includegraphics[height=4.5cm]{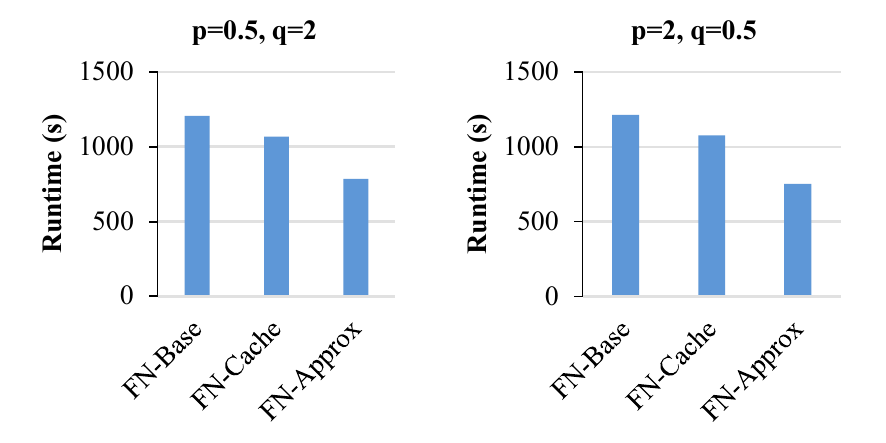}
  \vspace{-2mm}
  \\
  \mbox{(c) WeC-24}
  \\
  \end{array}
  \vspace{-0.05in}
$}
  \caption{Execution time of Fast-Node2Vec solutions for WeC-K graphs.}
  \label{fig:wec}
  \vspace{-0.1in}
\end{figure}

\begin{figure}[t]
\centering{$
  \begin{array}{c}
  \hspace{-0.15in}
  \includegraphics[width=8.9cm]{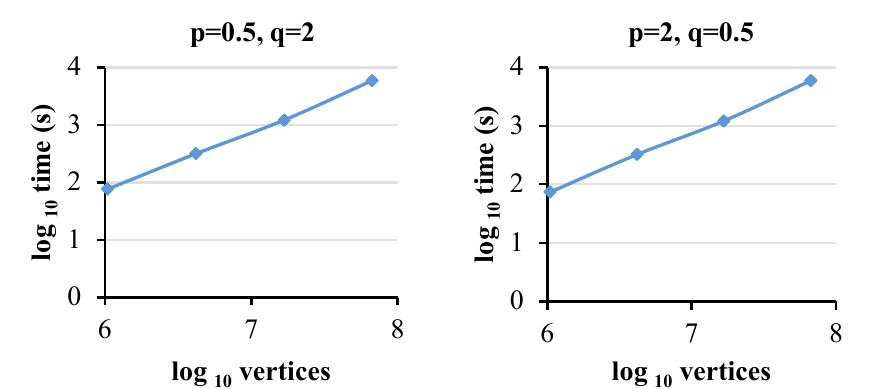}
  \end{array}
  \vspace{-0.1in}
$}
  \caption{Scalability of FN-Base on WeC-K.}
  \label{fig:wec-scale}
  \vspace{-0.1in}
\end{figure}

Figure~\ref{fig:ER} shows the scalability of FN-Base and C-Node2Vec on
the ER-K graphs varying the number of vertices from $2^{20}$ (about 1
million) to $2^{30}$ (about 1 billion).  The two figures show the
results for two sets of Node2Vec $(p,q)$ parameters.  Both the X-axis
and the Y-axis are in the log-scale.

From Figure~\ref{fig:ER}, we see that as the graph size increases,
C-Node2Vec scales linearly.  However, it runs out of memory for ER-K
graphs, where $K \geq 26$.  In comparison, our Fast-Node2Vec solution
scales linearly while the number of vertices increases from 1 million
to 1 billion.  FN-Base computes Node2Vec random walks on the largest
ER-K graph, i.e. ER-30, in about 2.3 hours.  This is quite reasonable
on a mid-sized cluster of machines.

Note that in an ER-K graph, the average degree is 10, and the degree
distribution is uniform.  Therefore, the optimization techniques for
popular vertices (including FN-Cache and FN-Approx) are not necessary.

\subsection{\hspace{-3mm}Efficiency and Scalability on WeC-K Graphs}
\label{subsec:WeC}

In this subsection, we study the efficiency and scalability of
Fast-NodeVec solutions on the WeC-K graphs.  Unlike the ER-K graphs,
the degree distribution in the WeC-K graphs are not uniform.  As shown
in Table~\ref{tab:datasets}, the maximum vertex degree is much (about
10--27 times) larger than the average vertex degree.  Therefore, we
expect the optimization techniques of FN-Cache and FN-Approx to be
beneficial.

Figure~\ref{fig:wec} shows the execution times of FN-Base, FN-Cache,
and FN-Approx for WeC-K graphs and for two sets of Node2Vec $(p,q)$
parameters.  We see that FN-Cache achieves a factor of 1.03--1.13x
improvements over FN-Base, and FN-Approx achieves a factor of
1.21--1.54x improvements over FN-Base.  This confirms our expectation.
In the next subsection, we will further study the relationship between
the skewness of the graphs and the impact of FN-Cache and FN-Approx on
execution time improvements.


Figure~\ref{fig:wec-scale} shows the scalability of FN-Base on WeC-K
graphs.  We see that FN-Base scales linearly while the number of
vertices increases from $2^{20}$ to $2^{26}$.

\subsection{In-Depth Analysis Using Skew-K Graphs}
\label{subsec:Skew}

\begin{figure}[t]
\centering{$
  \begin{array}{cc}
  \hspace{-5mm}
  \includegraphics[width=4.5cm]{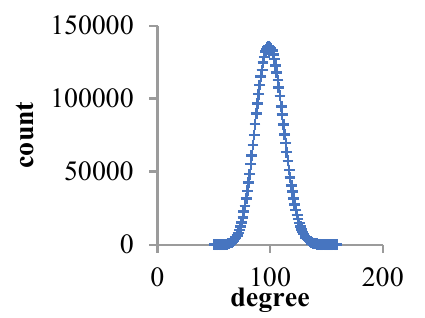} &
  \hspace{-5mm}
  \includegraphics[width=4.5cm]{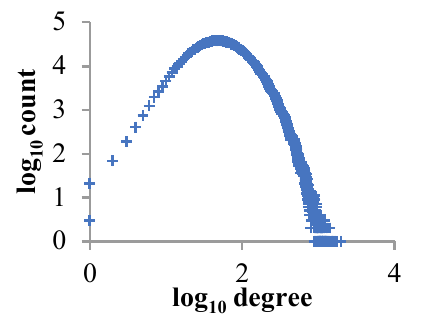}
  \\
  \mbox{(a) Skew-1 (Uniform)} &
  \mbox{(b) Skew-1.78 (WeC-22)}
  \vspace{0.05in}
  \\
  \hspace{-5mm}
  \includegraphics[width=4.5cm]{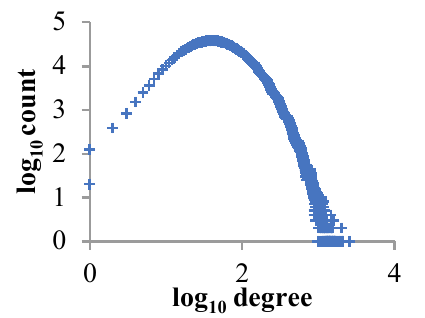} &
  \hspace{-5mm}
  \includegraphics[width=4.5cm]{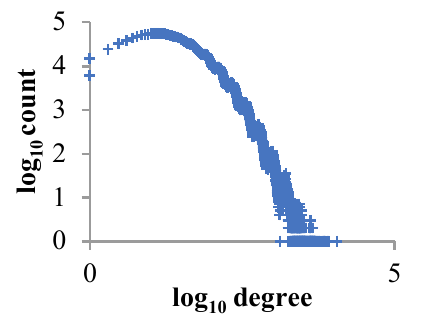}
  \\
  \mbox{(c) Skew-2} &
  \mbox{(d) Skew-3}
  \vspace{0.05in}
  \\
  \hspace{-5mm}
  \includegraphics[width=4.5cm]{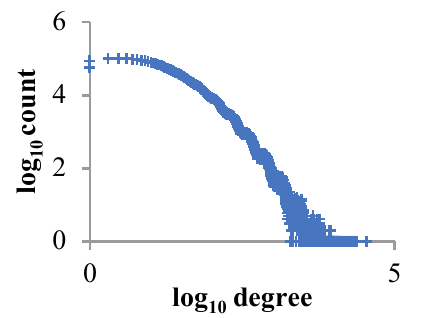} &
  \hspace{-5mm}
  \includegraphics[width=4.5cm]{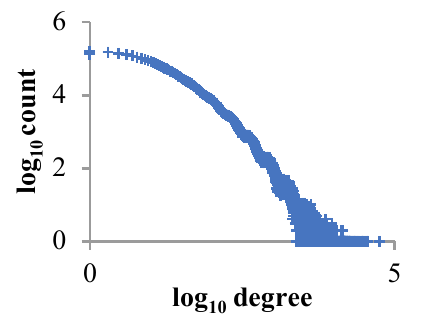}
  \\
  \mbox{(e) Skew-4} &
  \mbox{(f) Skew-5}
  \end{array}
  \vspace{-0.05in}
$}
  \caption{Vertex degree distribution of Skew-K.}
  \label{fig:skew-power-law}
  \vspace{-0.1in}
\end{figure}

\begin{figure}[t]
\centering{$
  \begin{array}{c}
  \hspace{-5mm}
  \includegraphics[height=4.5cm]{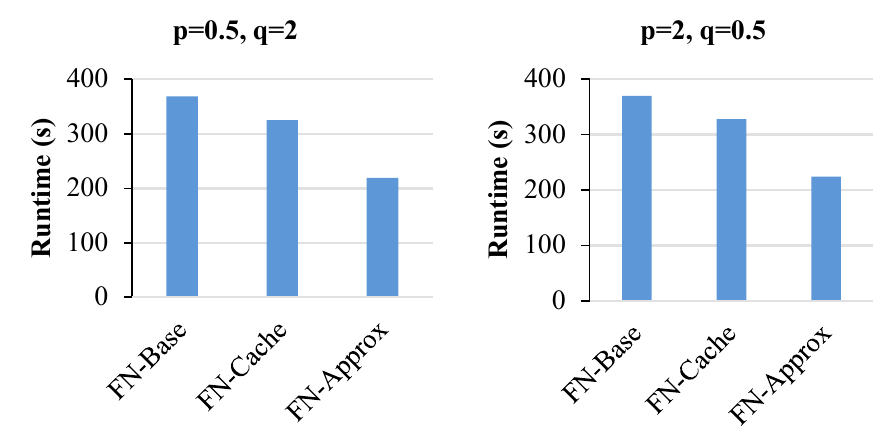} 
  \vspace{-3mm}
  \\
  \mbox{(a) Skew-2}
  \\
  \hspace{-5mm}
  \includegraphics[height=4.5cm]{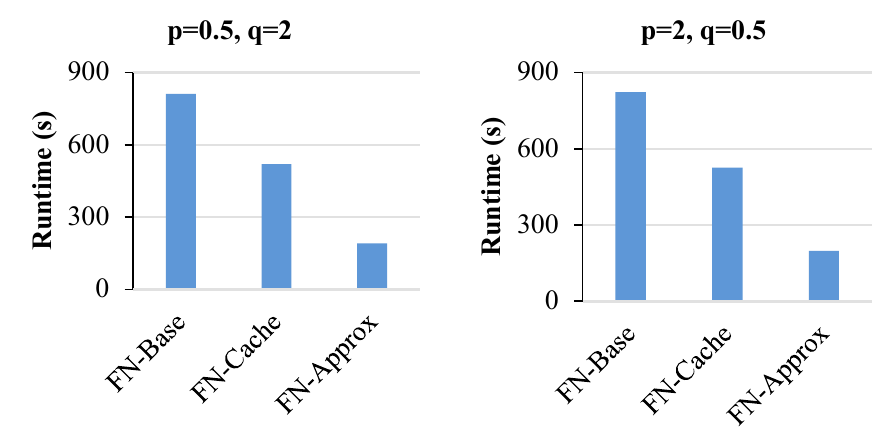}
  \vspace{-3mm}
  \\
  \mbox{(b) Skew-3}
  \\
  \hspace{-5mm}
  \includegraphics[height=4.5cm]{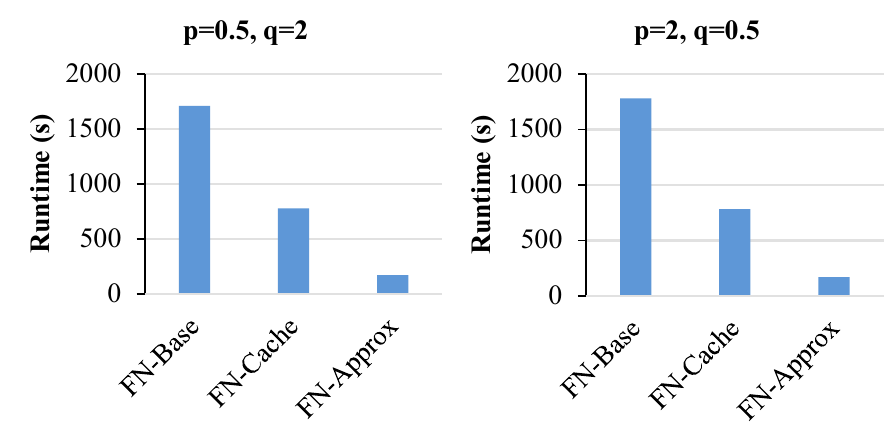}
  \vspace{-3mm}
  \\
  \mbox{(c) Skew-4}
  \\
  \hspace{-5mm}
  \includegraphics[height=4.5cm]{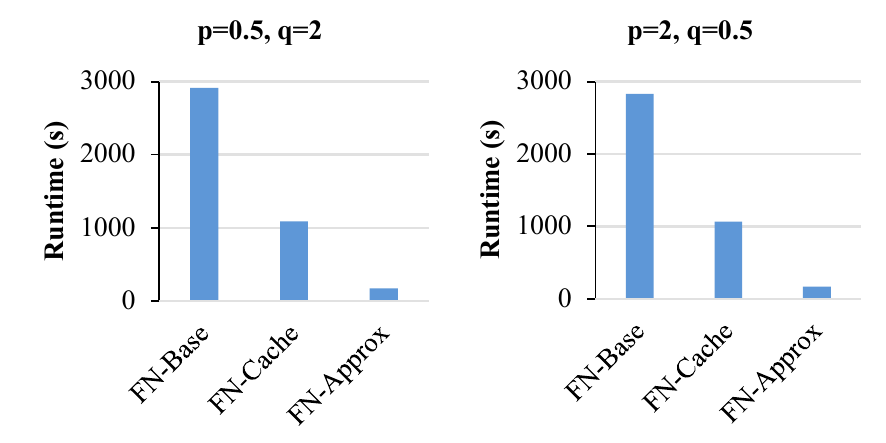}
  \vspace{-3mm}
  \\
  \mbox{(d) Skew-5}
  \\
  \end{array}
  \vspace{-0.05in}
$}
  \caption{Execution time of Fast-Node2Vec solutions for Skew-K, where
$K=2,\cdots,5$.}
  \label{fig:skew}
  \vspace{-0.1in}
\end{figure}

\begin{figure}[t]
\centering{$
  \begin{array}{cc}
  \hspace{-5mm}
  \includegraphics[height=4.0cm]{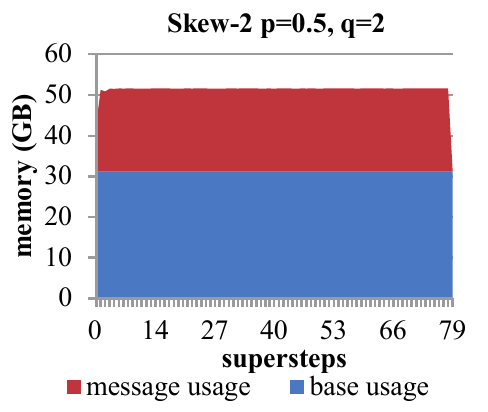}&
  \hspace{-7mm}
  \includegraphics[height=4.0cm]{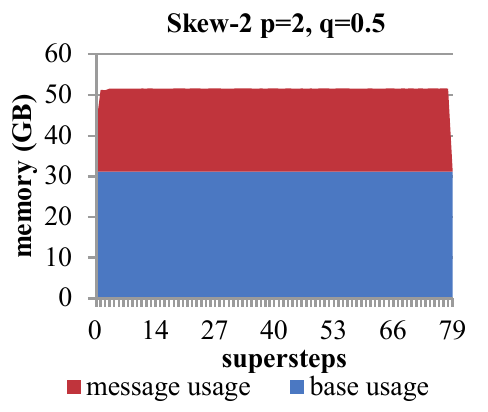}
  \\
  \hspace{-5mm}
  \includegraphics[height=4.0cm]{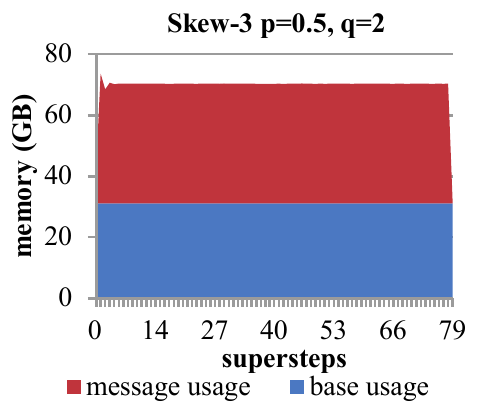}&
  \hspace{-7mm}
  \includegraphics[height=4.0cm]{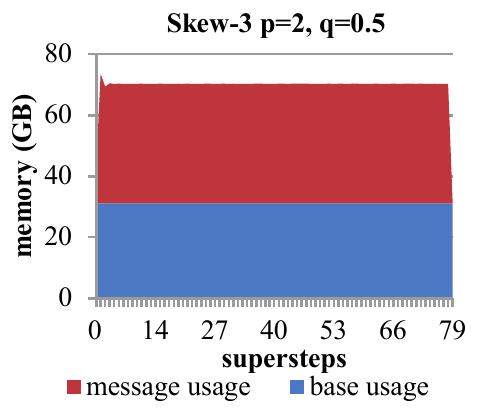}
  \\
  \hspace{-5mm}
  \includegraphics[height=4.0cm]{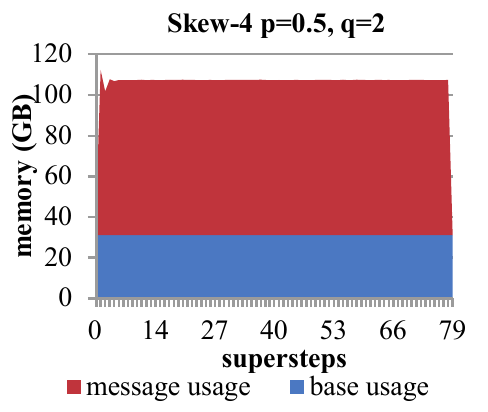}&
  \hspace{-7mm}
  \includegraphics[height=4.0cm]{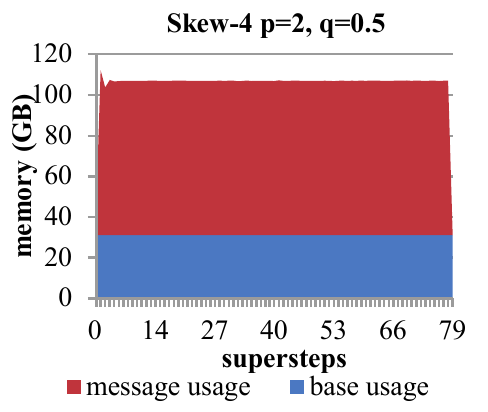}
  \\
  \hspace{-5mm}
  \includegraphics[height=4.0cm]{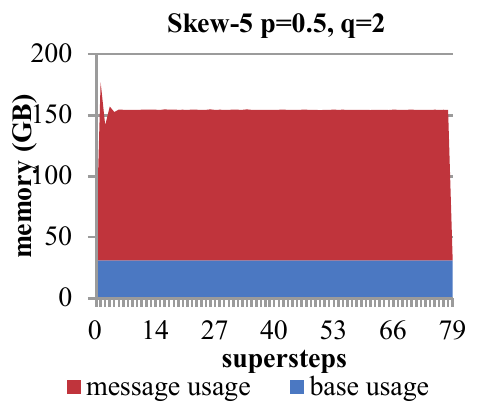}&
  \hspace{-7mm}
  \includegraphics[height=4.0cm]{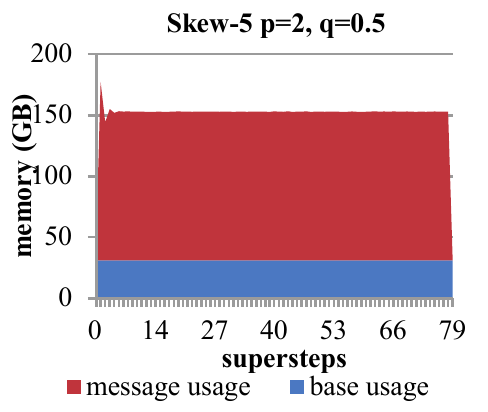}
  \end{array}
  \vspace{-0.05in}
$}
  \caption{Memory space consumption of FN-Base for Skew-K graphs, where
$K=2,\cdots,5$.}
  \label{fig:skew-mem}
  \vspace{-0.1in}
\end{figure}

Finally, we use the Skew-K graphs to analyze the relationship between
graph characteristics and the benefits of our proposed optimization
techniques.

Figure~\ref{fig:skew-power-law} shows the vertex degree distribution
of the Skew-K graphs.  We see that the vertex degree distribution of
Skew-1 is essentially a guassian distribution centered at 100.  This
is because the edges in Skew-1 are uniformly randomly generated.  When
$S>1$, the vertex degree distribution of Skew-S can be seen as a
combination of the guassian distribution (which is the arc shape in
the middle of the figure) and the power-law distribution (which is the
linear shape in the right part of the figure).  As $S$ grows larger
and larger, the distribution becomes closer and closer to the
power-law distribution.  Note that WeC-22 in Figure~\ref{fig:wec}(b)
is a Skew-S graph, where $S=1.78$.

Figure~\ref{fig:skew}(a)--(d) report the execution times of FN-Base,
FN-Cache, and FN-Approx for Skew-2, Skew-3, Skew-4, and Skew-5 with
two sets of Node2Vec $(p,q)$ parameters.  We make two interesting
observations from the figure:
\begin{list}{(\arabic{enumi})}{\setlength{\leftmargin}{5mm}\setlength{\itemindent}{0mm}\setlength{\topsep}{0.5mm}\setlength{\itemsep}{1mm}\setlength{\parsep}{0.5mm}}

\setcounter{enumi}{0}

\addtocounter{enumi}{1}
\item \emph{As $S$ increases, it takes longer time to compute Node2Vec
random walks}.  This can be clearly seen by the Y-axis labels.  In
particular, when $p=0.5, q=2$, FN-Base takes 314.5, 369.3, 811.1,
1710.4, and 2913.4 seconds for Skew-1.78, Skew-2, Skew-3, Skew-4, and
Skew-5, respectively.

\item \emph{As $S$ increases, the benefits of FN-Cache and FN-Approx
over FN-Base become larger}.  As $S$ grows from 1.78 to 5, the speedup
of FN-Cache over FN-Base grows from 1.04x to 2.68x when $p=0.5, q=2$,
and from 1.09x to 2.66x when $p=2, q=0.5$.  The speedup of FN-Approx
over FN-Base grows from 1.45x to 17.2x when $p=0.5, q=2$, and from
1.41x to 17.1x when $p=2, q=0.5$.

\end{list}

To better understand the results, we study the memory space
consumption of FN-Base for processing the Skew-K graphs, as shown in
Figure~\ref{fig:skew-mem}.   We see that as $S$ increases, the memory space
consumed by messages constitutes an increasingly larger portion of the
total space used.

Combining the results in Figure~\ref{fig:skew-power-law},
Figure~\ref{fig:skew}, and Figure~\ref{fig:skew-mem}, we see that as
$S$ increases, the vertex degree distribution is more and more skewed.
A greater many vertices have large numbers of neighbors.  Vertices
with larger degrees also tend to be sampled more frequently in the
random walks.  This leads to larger memory space allocated for
processing NEIG messages.  Consequently, \emph{optimizations for
popular vertices, including FN-Cache and FN-Approx, become more
effective when $S$ is larger}.

\section{Conclusion}
\label{sec:conclusion}

Node2Vec is a state-of-the-art feature learning method that generates
high-quality vector representations for the purpose of employing classical
machine learning methods in graph analysis.  However, we find that existing
Node2Vec implementations have significant difficulties in supporting
large-scale graphs.  The C++ and Python reference implementations of Node2Vec
are limited by the resource of a single machine.  An existing Node2Vec solution
on the Spark big data platform has poor result quality and incurs large
run-time overhead.  In this paper, we aim to efficiently support Node2Vec on
graphs with billions of vertices using a mid-sized cluster of machines. 

We propose and evaluate Fast-Node2Vec, a family of efficient Node2Vec random
walk algorithms.   Fast-Node2Vec employs GraphLite, a Pregel-like graph
computation framework, in order to avoid the overhead of read-only RDDs and
I/O-intensive shuffle operations in Spark.  It computes the transition
probabilities of Node2Vec random walks on demand, thereby reducing the memory
space required for storing all the transition probabilities, which is often
much larger than the total memory size in the mid-sized cluster for large-scale
graphs as shown in our evaluation.  Moreover, we also propose and evaluate a
set of techniques (e.g., FN-Cache and FN-Approx) to improve the efficiency of
handling large-degree verticies.  Experimental results show that Compared to
Spark-Node2Vec, fast-Node2Vec solutions achieve 7.7--122x speedups speedups.
Compared to the baseline Fast-Node2Vec, FN-Cache and FN-Approx achieve up to
2.68x and 17.2x speedups.

In conclusion, our proposed Fast-Node2Vec solutions can successfully support
Node2Vec computation on graphs with billions of vertices on a 12-node machine
cluster.  This means that researchers with moderate computing resources can
exploit Node2Vec for employing classical machine learning algorithms on
large-scale graphs.

\bibliographystyle{abbrv}
\bibliography{graph-ref}


\end{document}